\documentclass[aps,prd,10pt,twocolumn,nobibnotes,showpacs,showkeys,
superscriptaddress,nofootinbib]{revtex4-1}
\usepackage[dvips]{graphicx,graphics,color,epsfig}
\usepackage{amsmath,amsfonts,amssymb}
\usepackage{natbib}
\usepackage{verbatim}
\usepackage{latexsym}
\usepackage{dcolumn}
\usepackage{bm}
\begin{document}

\title{ENHANCEMENT OF THE STERILE NEUTRINOS YIELD AT HIGH MATTER DENSITY AND
AT INCREASING THE MEDIUM NEUTRONIZATION}

\author{\firstname{V.~V.}~\surname{Khruschov}}
\email{khruschov\_vv@nrcki.ru}
\affiliation{National Research Center ``Kurchatov Institute'',
Kurchatov~place~1, 123182~Moscow, Russia}

\author{\firstname{A.~V.}~\surname{Yudin}}
\email{yudin@itep.ru}
\affiliation{Institute for Theoretical and Experimental Physics,
ul. Bolshaya Cheremushkinskaya 25, Moscow, 117218 Russia}

\author{\firstname{D.~K.}~\surname{Nadyozhin}}
\email{nadezhin@itep.ru}
\affiliation{Institute for Theoretical and Experimental Physics,
ul. Bolshaya Cheremushkinskaya 25, Moscow, 117218 Russia}
\affiliation{National Research Center ``Kurchatov Institute'',
Kurchatov~place~1, 123182~Moscow, Russia}

\author{\firstname{S.~V.}~\surname{Fomichev}}
\email{fomichev\_sv@nrcki.ru}
\affiliation{National Research Center ``Kurchatov Institute'',
Kurchatov~place~1, 123182~Moscow, Russia}
\affiliation{Moscow Institute of Physics and Technology, 141700 Dolgoprudny,
Moscow Region, Russia}

\begin{abstract}
The relative yields of active and sterile neutrinos in the matter with a high
density and different degree of neutronization are calculated. A significant
increase in the proportion of sterile neutrinos produced in superdense matter
when degree of neutronization approaches the value of two is found.
The results obtained can be used in the calculations of the neutrino fluxes
for media with a high density and different neutronization degrees in
astrophysical processes such as the formation of protoneutron core of a
supernova.
\end{abstract}

\pacs{14.60.Pq, 14.60.St, 12.10.Kt, 12.90.+b, 26.30.Jk}

\keywords{active and sterile neutrinos, neutrino oscillations, interaction of
neutrino with matter, ratio of number of neutrons to number of protons,
protoneutron core of supernova}

\maketitle

\section*{INTRODUCTION}
\noindent
At gravitational collapse of supernovae main part of the released energy is
passed away by powerful flows of neutrino. Neutrinos of various types
(flavours) are born both due to a large number of the processes with
participation of nucleons and nuclei of stellar matter and due to processes
of neutrino oscillations, i.e. to transitions of one type of neutrino to
another. The specific feature of processes of oscillations of known types of
neutrino (electron, muon and tau) in matter is their dependence on
distribution of the electron density. At the nonzero electron density the
oscillation characteristics of neutrino are changed in comparison with the
oscillation characteristics of neutrino in vacuum, and at certain
relationships between the electron density, differences of squares of neutrino
masses, angles of mixing and neutrino energy the so-called
Mikheev--Smirnov--Wolfenstein (MSW) resonances arise. In those areas of star
matter, where conditions of the MSW-resonances are satisfied, an
intensification of transitions of one type of neutrino to another occurs, even
if the initial vacuum mixing between different types of neutrino is
insignificant.

The accounting of strengthening of transitions of one neutrino type to another
due to the MSW-resonance in solar matter allowed to solve a problem of
deficiency of the solar electron neutrino and to determine the difference of
squares of neutrino masses $\Delta m^2_{21}=m^2_{2}-m^2_{1}$ and the neutrino
mixing angle $\theta_{12}$. On the basis of data on oscillations of
atmospheric, reactor and accelerator neutrinos other oscillation
characteristics of neutrino were also determined (see Olive et al., 2014).
Now most exact values of vacuum oscillation characteristics of neutrino in
the limits of deviations up to $1\sigma$, where $\sigma$ is standard
uncertainty measurements, are obtained in a number of papers. We present the
values of the oscillation characteristics from the paper of Gonzalez-Garcia
et al. (2014) for the standard parametrization of a mixing matrix:
\begin{subequations}
\begin{equation}
\sin^2\theta_{12}=0.304^{+0.013}_{-0.012}\,,\label{eq1a}\tag{1a}
\end{equation}
\begin{equation}
\sin^2\theta_{23}=\left\{\begin{array}{lr}{\rm NH}:&0.452^{+0.052}_{-0.028}\\
{\rm IH}:&0.579^{+0.025}_{-0.037}
\end{array}\right.\!,\label{eq1b}\tag{1b}
\end{equation}
\begin{equation}
\sin^2\theta_{13}=\left\{\begin{array}{lr}{\rm NH}:&0.0218^{+0.0010}_{-0.0010}
\\
{\rm IH}:&0.0219^{+0.0011}_{-0.0010}\end{array}\right.\!,\label{eq1c}\tag{1c}
\end{equation}
\begin{equation}
\Delta m_{21}^2/10^{-5}{\rm {\text{eV}}}^2=7.50^{+0.19}_{-0.17}\,,\label{eq1d}
\tag{1d}
\end{equation}
\begin{equation}
\Delta m_{31}^2/10^{-3}{\rm {\text{eV}}}^2=
\left\{\begin{array}{lr}{\rm NH}:&2.457^{+0.047}_{-0.047}\\
{\rm IH}:&-2.449^{+0.048}_{-0.047}\end{array}\right.\!.\label{eq1e}\tag{1e}
\end{equation}
\label{dat}
\end{subequations}
As only the absolute value of oscillation mass characteristic
$\Delta m^2_{31}$ is known, the absolute values of neutrino masses can be
ordered in two ways: $a) \; m_1<m_2<m_3$ and $b) \; m_3<m_1<m_2$, that is,
it can be realized, as the saying goes, either normal hierarchy
(NH, case {\it a}), or inverted hierarchy (IH, case {\it b}) of the neutrino
mass spectrum.

Along with the given values of oscillation characteristics of neutrinos, for a
long time there are evidences of anomalies of neutrino flows in various
processes occurring on Earth, which can not be explained by oscillations of
only active, i.e. electron-, muon- and tau-neutrino and antineutrino.
LSND/MiniBooNE, reactor and gallium anomalies belong to such anomalies (see
Abazajian et al., 2012; Kopp, et al., 2014), which could be explained by
existence of one or two additional neutrinos noninteracting within the
Standard Model (SM) with other particles. Such neutrinos were named as
sterile neutrinos. The characteristic mass scale of the sterile neutrinos,
which is responsible for the description of the anomalies noted above is 1~eV.
However, it should be noted that recently obtained observational astrophysical
data pertaining to formation of galaxies and their clusters can be explained
by existence of sterile neutrinos with the mass of the order of 1~KeV or
above, and such sterile neutrinos are the candidates for cold dark matter
particles (see Dodelson and Widrow, 1994; Kusenko, 2009). More details about
possible existence of sterile neutrinos and their characteristics can be found
in numerous papers (see, for example, Liao, 2006; Abazajian et al., 2012;
Bellini et al., 2013; Conrad et al., 2013; An et al., 2014; Kopp et al.,
2014).

Of great interest are the models with three active and three sterile neutrinos
(see Bhupal Dev and Pilaftsis, 2012; Duerr et al., 2013; Conrad et al., 2013;
Rajpoot et al., 2013; Khruschov, 2013; Zysina et al., 2014; Khru\-schov and
Fomichev, 2015), which can be included in the grand unification theories (GUT)
keeping up the left-right symmetry. In the papers by Zysina et al. (2014) and
Khruschov and Fomichev (2015) the estimates of masses of three active and
three sterile neutrinos were obtained and both the appearance and survival
probabilities of active and sterile neutrinos in the Sun with accounting of
the MSW-resonances were calculated. Although in the models with three active
and three sterile neutrinos the difficulties exist, which are connected with
consistency of a number of effective neutrino types, as well as the sum of
their masses with the data of cosmological observations related to primary
nucleosynthesis, CMB anisotropy and the observed large-scale structure of the
universe (see Komatsu, et al., 2011; Ade, et al., 2013), such consistency
depends on the cosmological model used. Actually, it is possible to bypass a
direct link between the number of additional relativistic degrees of freedom
and the number of sterile neutrinos by using more general cosmological models
(see Ho and Scherrer, 2013; Gorbunov, 2014; and references therein). In this
paper we consider for the sake of generality a model with three sterile
neutrinos, in which the phenomenological parameters are chosen in order to
demonstrate the effect of increase of the sterile neutrino yield at certain
conditions. The taken parameter values do not contradict to the existing
experimental limitations. There is a simple way to suppress the production and
thermalization of sterile neutrinos in the early universe, which can be
applied in our model, too. This is reducing or even vanishing the mixing
parameters between the active neutrinos and one, two or three sterile
neutrinos and disturbing all or some of sterile neutrinos from an
thermodynamic equilibrium at early stages of formation of the Universe, That
is, the model admits a reduction to two or even one sterile neutrino.

In the current paper, the neutrino flavour composition modification due to
coherent scattering of a neutrino on both electrons and neutrons is
considered. The accounting of neutrons does not lead to change of oscillation
characteristics, if only the active neutrinos are considered. But if a
contribution of the sterile neutrinos is taken into account, the influence of
neutron density becomes noticeable. Moreover, it is shown in the current paper
that, when the ratio of a number of neutrons to a number of protons in the
matter is close to two, there is a considerable enhancement of the sterile
neutrino yield. Such enhancement arises at large or super-large values of
density ($>10^7$~g/cm$^3$). Therefore, this effect can be of importance only
in the astrophysical conditions, for example, at formation of a protoneutron
core of a supernova. In the models with participation of the sterile neutrinos
this effect is additional to the MSW-effect and can lead to new consequences
at supernovae explosions. In spite of the fact that influence of the sterile
neutrinos on the processes in supernovae were considered in many papers (see,
for example, Hidaka and Fuller, 2006 and 2007; Tamborra, et al., 2012; Wu,
et al., 2014; Warren, et al., 2014), the effect of enhancement of the sterile
neutrino yield at the ratio of the number of neutrons to the number of protons
in the medium given as $\eta=N_n/N_p\approx 2$ was not noticed before.

\section*{THE (3+1+2)-MODEL OF ACTIVE AND STERILE NEUTRINOS}
\noindent
We will give below the basic principles of the (3+1+2)-model with three active
and three sterile neutrinos investigated in the papers by Zysina et al. (2014)
and Khruschov and Fomichev (2015). Within the (3+1+2)-model, a neutrino of
certain types of flavour $\{\nu_f\}$, both active and sterile, are the mix of
massive neutrinos $\{\nu_m\}$ having a certain masses. The neutrino masses
are given by a set $\{m\}=\{m_i,m_{i'}\}$, where $i=1,2,3$, $i'=1',2',3'$,
and the masses $\{m_i\}$ settle down in a direct order as $m_1,m_2,m_3$,
whereas the masses $\{m_{i'}\}$ settle down in the inverse order as
$m_{3'},m_{2'},m_{1'}$. The full set $\{\nu_f\}=\{\nu_a,\nu_s\}$ of the
flavour neutrino consists of the known active neutrinos $\{\nu_a\}$, i.e.
$e$-, $\mu$-, and $\tau$-neutrino, and three hypothetical sterile neutrinos
$\{\nu_s\}$, which we will distinguish by indexes $x$, $y$ and $z$. The
generalized $6\times 6$ mixing matrix $\tilde{U}$ can be presented by means of
$3\times3$ matrices $S$, $T$, $V$ and $W$ as follows:
\begin{equation}
\left(\begin{array}{c}
\nu_{\alpha}\\
\nu_s
\end{array}
\right)=
\tilde{U}
\left(\begin{array}{c}
\nu_{i}\\
\nu_{i'}
\end{array}
\right)
\equiv
\left(\begin{array}{cc}
S&T\\
V&W
\end{array}
\right)\left(\begin{array}{c}
\nu_{i}\\
\nu_{i'}
\end{array}
\right).
\label{eq/2}
\end{equation}

Hereafter, the particular form of matrix $\tilde{U}$ will be used taking into
account the smallness of parameters of mixing between the active and the
sterile neutrinos, and also assuming that the mixing of the sterile states
$\{\nu_{i'}\}$ can be neglected ($W=1$ is the unit matrix). Then, in condition
of conservation of the $CP$-invariance in the lepton sector, the matrices $S$,
$T$ and $V$ can be written down in the following form:
\begin{subequations}
\begin{align}
&S = U_{PMNS} + \Delta U_{PMNS},\label{eq3a}\tag{3a}\\
&T = b, \quad V = -b^T U_{PMNS},\label{eq3b}\tag{3b}
\end{align}
\label{eq/3}
\end{subequations}
where $U_{PMNS}$ is a mixing matrix for the active neutrinos, i.e. the
Pontekorvo--Maki--Nakagawa--Sakata matrix. The contributions from the matrix
elements of $\Delta U_{PMNS}$ are actually allowed for by the experimental
uncertainties of the matrix elements of $U_{PMNS}$. In what follows we will
choose the case of inverted hierarchy ($IH$) of the active neutrinos mass
spectrum, which is preferable for the $\nu_{\mu}-\nu_{\tau}$-symmetry of a
neutrino mass matrix, and also for the detailed description of the survival
probability of the solar electron neutrinos in the energy range higher than
2~MeV (see Khrushchov and Fomichev, 2015).

For the $IH$-case, the matrix $b$ can be given as follows
\begin{equation}
b_{IH}=
\left(\begin{array}{lcr}
\gamma&\gamma'&\gamma'\\
\beta&\beta'&\beta' \\
\alpha&\alpha'&\alpha'
\end{array}
\right),
\label{eq/4}
\end{equation}
where the parameters $\alpha,\beta,\gamma,\alpha',\beta',\gamma'$
should be in the range from zero up to $0.2$ in absolute values.

The specific feature of the (3+1+2)-model is the mass spectrum of the sterile
neutrinos, one of which is rather heavy and can in principle be with a mass
from 0.5~eV up to several keV and above, but two others are light with masses
about 2~meV (see Zysina et al., 2014). In the current paper we will use
the following values of the mass parameters in addition to experimental data
given above in Eqs.~(\ref {dat}), the neutrino masses being given in~eV:
\begin{equation}
m_1=0.0496,\quad m_2=0.0504, \quad m_3=0.002,
\label{eq/5}
\end{equation}
\begin{equation}
m_1'=0.002, \quad m_2'=0.0022, \quad m_3'=0.46,
\label{eq/6}
\end{equation}
\begin{equation}
\beta=\gamma=0.1,\quad \beta'=\gamma'=0,
\label{eq/7}
\end{equation}
\begin{equation}
\alpha=0,\quad \alpha'=0.15.
\label{eq/8}
\end{equation}
The value of $m_3'=0.46$~eV corresponds to the mass of the heavy sterile
neutrino given in the paper by Sinev (2013). Notice that the  values of
parameters given above are phenomenological, i.e., they are chosen taking into
account the available experimental restrictions. The choice of concrete values
of parameters is necessary both for the investigation of the (3+1+2)-model and
carrying out the numerical calculations, and also for the demonstration of new
effect of enhancement of the sterile neutrino yield in matter
with high values of density and at neutronization degree $\eta=N_n/N_p$, that
is the ratio of a number of neutrons to a number of protons, close or equal to
two.

\section*{ENHANCEMENT OF THE STERILE NEUTRINO YIELD}
\begin{figure}[tbp]
\centering
\includegraphics[width=0.48\textwidth]{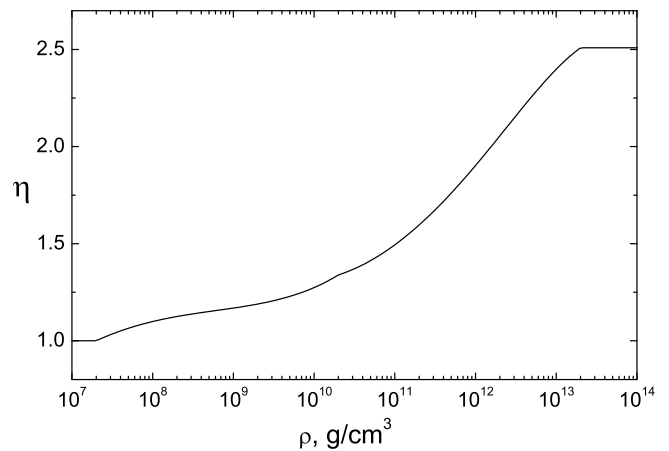}
\caption{The dependence of the neutronization coefficient $\eta$ on
density $\rho$, which is specific for a supernova core at conditions of a
gravitational collapse.}
\label{Figure1}
\end{figure}

\noindent
Let us write down the equation for the probability amplitudes of a neutrino
with certain flavours, which propagates in the matter, in the form given
in the paper by Khruschov and Fomichev (2015):
\begin{align}
&i\partial _{r}\left(\begin{array}{c}
a_{\alpha} \\
a_{s}\end{array}\right)=\nonumber\\
&=\left[\frac{\tilde{\Delta}_{m^2}}{2E}
+\sqrt{2}G_{F}\left(\begin{array}{cc}
\tilde{N}_{e}(r) & 0 \\
0 & \tilde{N}_{n}(r)/2\end{array}\right)\right]
\left(\begin{array}{c} a_{\alpha}\\
a_{s}\end{array}\right).
\label{ur}
\end{align}
Here the matrix $\tilde{\Delta}_{m^2}$ is defined as
$\tilde{\Delta}_{m^2}=\tilde{U}\Delta_{m^2}\tilde{U}^T$,
$\Delta_{m^2}={\rm diag}\{m_{1}^{2}-m_{0}^{2},m_{2}^{2}-m_{0}^{2},m_{3}^{2}
-m_{0}^{2},m_{3'}^{2}-m_{0}^{2},m_{2'}^{2}-m_{0}^{2},m_{1'}^{2}-m_{0}^{2},\}$,
with $m_0$ being the smallest neutrino mass among $m_i$ and $m_{i'}$, and
$\tilde{N}_{e}(r)$ and $\tilde{N}_{n}(r)$ are $3\times3$ matrices defined as:
\begin{equation}
\tilde{N}_{e}(r)=\left(\begin{array}{ccc}
N_{e}(r) & 0 & 0 \\
0 & 0 & 0 \\
0 & 0 & 0 \end{array}\right),
\label{eq/10}
\end{equation}
\begin{equation}
\tilde{N}_{n}(r)=\left(\begin{array}{ccc}
N_{n}(r)& 0 & 0 \\
0 & N_{n}(r) & 0 \\
0 & 0 & N_{n}(r) \end{array}\right),
\label{eq/11}
\end{equation}
where $N_e(r)$ and $N_n(r)$ are the local electron and neutron densities,
respectively, in the star.\footnote{Note that the contribution due to
neutrino-neutrino interaction dependent on the neutrino density is omitted in
the equation (\ref{ur}). The reason is that we consider the general problem of
the interaction of neutrinos with a dense medium, in which there is no large
density of neutrinos, at least initially. So, the equation in the form of
(\ref{ur}) can be applied to the formation of protoneutron core of the
supernova, if we consider only the processes occurring in the initial stage of
radiation of electron nonthermal neutrinos (see Fig. 5 in the paper of
Nadyozhin and Imshennik, 2005).} We will find out the specific features of
solutions of the equation (\ref {ur}) at certain values of $N_n$ and
$N_e\equiv N_p$, considering the electroneutral star medium.
Figure~\ref{Figure1} shows the behavior of the neutronization degree $\eta$ in
a star during the collapse right up to "bounce" of the core
(see Liebend\"{o}rfer, 2005). Each mass shell of the collapsing stellar core
increases its density $\rho$ by approximately similar way. Thus, at a collapse
stage $\eta=\eta(\rho)$. The value of $\eta=2$ corresponds to the density
$\rho\approx 2\times 10^{12}$~g/cm$^3$.

\begin{figure*}[tbp]
\centering
\includegraphics[width=0.99\textwidth]{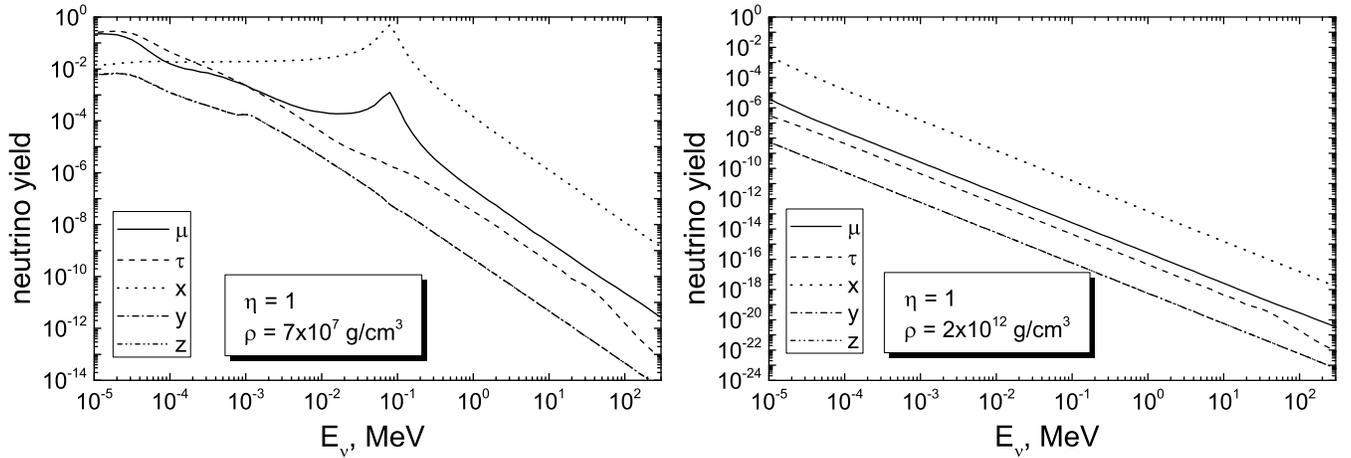}
\caption{The relative yields of various (not electron) flavours of the
neutrino versus the neutrino energy at $\eta=1$ at a full normalization of the
neutrino yield on the unity. The density $\rho=7\times10^{7}$~g/cm$^3$
(the left panel) and $\rho=2\times10^{12}$~g/cm$^3$ (the right panel).
$\Delta r=20$~km. The yields of sterile {\it y}\,- and {\it z}\,-neutrinos
practically coincide.}
\label{Figure2}
\end{figure*}
\begin{figure*}[tbp]
\centering
\includegraphics[width=0.99\textwidth]{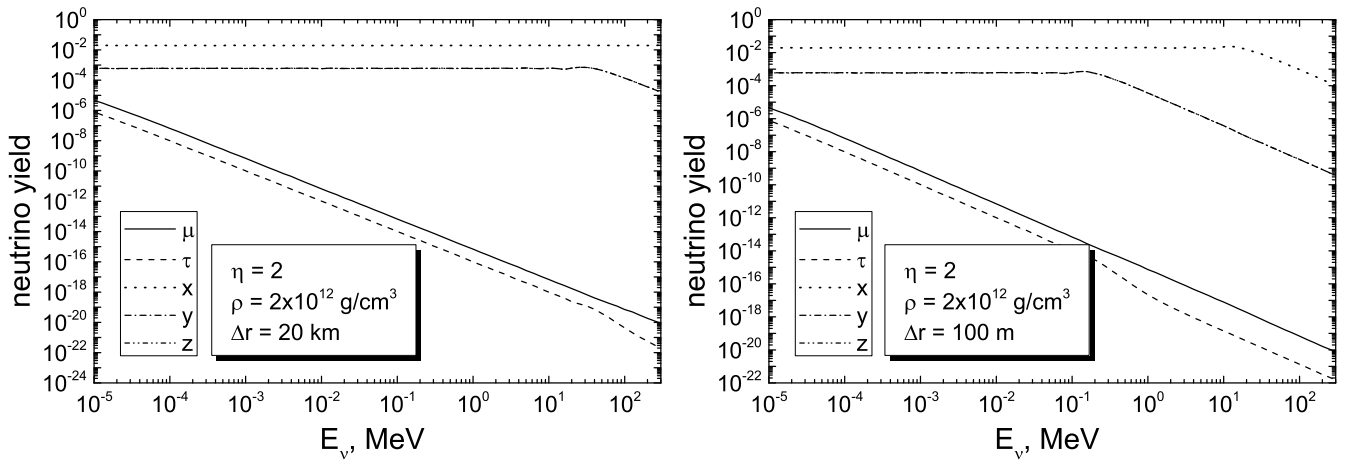}
\caption{The relative yields of various (not electron) flavours of the
neutrino versus the neutrino energy at $\eta=2$ and at the density
$\rho=2\times10^{12}$ g/cm$^3$, and also at two values of the spatial range
$\Delta r$, where solutions of equation (\ref {ur}) were obtained. The yields
of sterile {\it y}\,- and {\it z}\,- neutrinos practically coincide.}
\label{Figure3}
\end{figure*}
\begin{figure*}[tbp]
\centering
\includegraphics[width=0.99\textwidth]{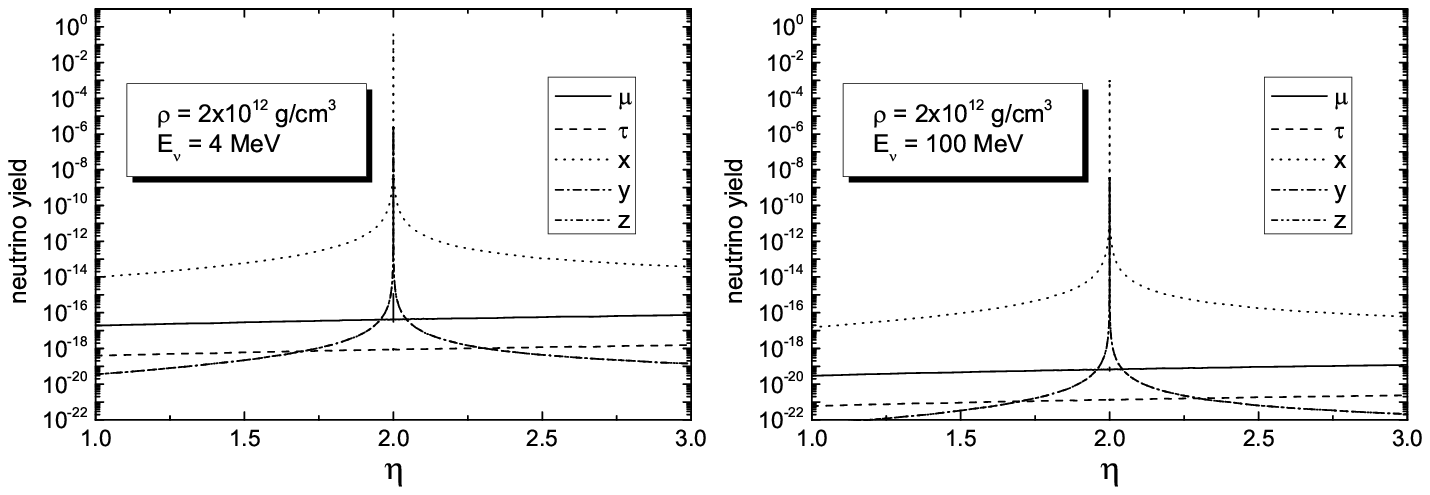}
\caption{The relative yields of various (not electron) flavours of the
neutrino versus the neutronization degree $\eta$ in the range of $\eta$-values
around $\eta=2$ at the density $\rho=2\times10^{12}$~g/cm$^3$. $E_\nu=4$~MeV
(the left panel) and $E_\nu=100$~MeV (the right panel). $\Delta r=100$~m. The
yields of sterile {\it y}\,- and {\it z}\,- neutrinos practically coincide.}
\label{Figure4}
\end{figure*}

Figure~\ref{Figure2} displays the results obtained for the relative average
yields of various (not electron) flavours of the neutrino (at a normalization
of the total yield of all flavours, including the electron one, on unity)
versus the neutrino energy $E_{\nu}$ at values of density
$\rho=7\times10^{7}$~g/cm$^3$ (the left panel) and
$\rho=2\times 10^{12}$~g/cm$^3$ (the right panel), for the neutronization
degree $\eta=1$. This value of $\eta$ is typical for the normal (equilibrium)
conditions of matter. For the results of Fig.~\ref{Figure2}, the
equation~(\ref{ur}) was solved with a constant density on a spatial scale of
the order of typical radius of the neutron star, exactly, on the scale
$\Delta r=20$~km. For obtaining the average yields of neutrinos the solutions
of this equation were averaged on the spatial scale of neutrino oscillations.
As concerns the initial conditions, it was assumed that only electron
neutrinos are created in the very beginning. In this Figure, as expected, the
relative yields of the non-electron neutrino flavours are small, and for
neutrino energies $E_{\nu}$ more than 1~MeV they do not exceed $10^{-5}$. On
the left panel in the range of $E_{\nu}$ from $10^{-2}$~MeV up to
$10^{-1}$~MeV, increasing of the yields of two flavours of neutrino connected
with the MSW-effect can be seen. The relative yields given on the right panel
for the non-electron neutrino flavours are small everywhere, and for the
$E_{\nu}$-range from 10~eV up to 0.1~GeV they do not exceed $10^{-10}$.

However, the situation sharply changes at approaching the matter
neutronization degree $\eta$ to value of two at large density.
Figure~\ref{Figure3} shows the relative yields of various non-electron
flavours of neutrino versus the neutrino energy at $\eta=2$ and at the same
density as on the right panel of Fig.~\ref{Figure2}. The results are given for
two different values of spatial scale $\Delta r$, where the
equation~(\ref {ur}) was solved. In both cases a visible increase of the
relative yields of sterile neutrinos of $x$\,-, $y$\,- and $z$-flavours can be
seen, and here their yields, unlike the MSW-resonances, poorly depend on the
energy of the neutrino. Note that the second case presented on the right panel
of Fig.~\ref{Figure3} with $\Delta r=100$~m is physically more preferable
owing to two factors: narrowness of the effective spatial range with
$\eta\approx 2$ in a collapsing star and rather small neutrino mean free path
at such density. Namely, the mean free path of neutrino in this case is equal
approximately to $10-100$~m.\footnote{Mean free path of neutrino in matter is
$l_\nu=1/n\sigma$, where $n\sim\rho/m_{\mathrm n}$ is the concentration of
nucleons, and interaction cross-section of neutrino is
$\sigma\sim\left(E_\nu/m_{\mathrm e}c^2\right)^2~10^{-44}$~cm$^2$, with
$m_\mathrm e$ the electron mass, i.e., for $\rho\sim10^{12}$~g/cm$^3$ and
$E_\nu\sim 100$~MeV $l_\nu$ is of the order of tens meters. The oscillation
length of sterile neutrino can also be estimated as proportional to
$E_{\nu}/\Delta m^2$, and for the neutrino energy $E_{\nu}=100$~MeV and for
the neutrino mass of the order of 1~eV it has the order of several tens meters
as well. Therefore, at $E_{\nu}\lesssim 100$~MeV the oscillation length is
much lesser than the neutrino mean free path, and it permits, in particular,
to use Eq.~(\ref{ur}) without allowance for the effect of coherence loss at
these neutrino energies.}

To clarify nature of the neutrino yield when $\eta$ approaches two at large
density, the relative average yields of various non-electron flavours of
neutrino are shown in Fig.~\ref {Figure4} (at a normalization of total yield
of all neutrino flavours, including the electron one, to unity). The left
panel of Fig.~\ref{Figure4} corresponds to density
$\rho=2\times 10^{12}$~g/cm$^3$ and neutrino energy $E_\nu=4$~MeV, while the
right panel corresponds to the same density and neutrino energy
$E_\nu=100$~MeV. It is seen that at the density $\rho=2\times10^{12}$~g/cm$^3$
the sharp enhancement of the relative yields of various flavours of sterile
neutrinos occurs at $\eta=2$, while the yields of active $\mu$- and
$\tau$-neutrino only hardly change at variation of $\eta$. Note that the
latter is connected with the chosen structure of parameters of a generalized
mixing matrix.

The resonance curves presented in Fig.~\ref{Figure4} were obtained at the
numerical calculations on the spatial scale $\Delta r=100$~m corresponding
approximately to the scale of the range of excess neutronization (closely to
two) in the collapsing star (see Fig.~\ref{Figure1}), and also it corresponds
to the scale of the order of the neutrino mean free path in such dense medium.
By these results, it is possible to determine the relation between the width
of the resonance curves over neutronization degree $\eta$ and the range of the
resonant enhancement of the sterile neutrino yield in the star.

Let us consider a question about the width of the resonant enhancement zone
of the neutrino oscillations in the star. In accordance with the calculations
given above, it corresponds to a narrow zone around $\eta\approx2$. The
hydrostatic equilibrium equation in the star reads
\begin{equation}
\frac{1}{\rho}\frac{d P}{d r}=-\frac{G m}{r^2},
\end{equation}
where $\rho$ and $P$ are the density and the pressure in the star, and $m$ and
$r$ are the mass and radial coordinates, respectively. For the pressure
gradient, taking into account approximate constancy of entropy in internal
areas of a star, we will obtain
\begin{equation}
\frac{d P}{dr}=\frac{d P}{d\rho}\frac{d\rho}{dr}\approx\gamma\frac{P}{\rho}
\frac{d\rho}{dr},
\end{equation}
where
\begin{equation}
\gamma\equiv\left(\frac{\partial\ln P}{\partial\ln\rho}\right)_S
\end{equation}
is the adiabatic index. Then passing to the finite differences
we obtain that
\begin{equation}
\triangle r \approx\gamma\frac{P}{\rho^2}\frac{r^2}{Gm}\triangle\rho=
\gamma\frac{P}{\rho}\frac{r^2}{Gm}\left[\frac{\triangle\eta}{\rho\frac{d\eta}
{d\rho}}\right].\label{dr_deta}
\end{equation}
The quantity before the square brackets in equation (\ref {dr_deta}) can be
roughly estimated from the calculations of gravitational collapse and the
equation of state of supernova matter (see, for example, Nadyozhin and Yudin
2004) by the characteristic value of $5\times10^5$~cm. The value of derivative
of neutronization degree $\eta$ at the density $\rho=2\times10^{12}$~g/cm$^3$
can be found from Fig.~\ref{Figure1}: $\rho\frac{d\eta}{d\rho}\approx0.22$.
Hence, the width of the resonance zone over $r$ in a star is
\begin{equation}
\triangle r\approx 2\times 10^{6}\ \triangle\eta~\mbox{[cm]},
\end{equation}
where $\triangle\eta$ is the resonance width over the neutronization degree,
which can be determined, for example, from Fig.~\ref{Figure4}. By virtue of
narrowness of the resonance zone and huge magnitude of the effect (of many
orders), a concrete value of the width depends on its definition. We consider,
for example, the case of the neutrino energy $E_\nu=100$~MeV (the right panel
of Fig.~\ref{Figure4}). Let us determine the resonance width as the zone, in
which the sterile {\it x}-neutrino yield falls by seven orders of magnitude
from its maximum value ($\sim10^{-3}$), reaching the values of the order of
$10^{-10}$ (after that it also falls by seven orders of magnitude, reaching
the smallest values of the order of $10^{-17}$ at the ends of the inspected
$\eta$\,-range, at $\eta=1$ or $\eta=3$). In this case we obtain
$\triangle\eta\approx0.003$ that corresponds to $\triangle r\approx 60$~m. All
the above with the same result is applicable also to the left panel of
Fig.~\ref{Figure4}, i.e. for neutrino energy $E_\nu=4$~MeV. The estimate
obtained shows the self-consistency of the carried out calculations of new
neutrino resonances for sterile neutrinos in superdense medium of neutron
stars.

\section*{CONCLUSION}
\noindent
The paper presents the effect of resonant enhancement of the sterile neutrino
yields at neutronization degree $\eta=2$, which is considered in the framework
of the model problem on the basis of equation (\ref{ur}). This effect, which
poorly depends on the neutrino energy at $E_{\nu}$ more than $10$~eV, can have
a significant impact on dynamics of gravitational collapse of a supernova star
and subsequent expulsion of its envelope, that we suppose to investigate in
further calculations. Note that in our model problem, the main purpose of
which was to demonstrate the existence and importance of the effect of
enhancement of the sterile neutrino yield at $\eta=2$, it is possible as a
first approximation to neglect the effect of the loss of coherence due to weak
scattering in the equation (\ref{ur}) at neutrino energies not exceeding
$\sim 100$~MeV. The effect of loss of coherence due to the
weak scattering in certain conditions where it may be substantial, as well as
other possible specific features of the actual situation in the collapsing
star will be taken into account quantitatively in the further development of
the model.

The effect of resonant enhancement of sterile neutrino yield at a
neutronization degree of the medium equal to two still remains in a reduced
model with only a single sterile neutrino. This effect takes place for each of
sterile neutrinos independently of the presence of other sterile neutrinos. Up
to now, the question about the number of sterile neutrinos has not been
completely resolved, both theoretically and experimentally. At present, a
large number of international laboratory experiments aimed at searching for
the sterile neutrinos are planned to be implemented. The conclusion following
from the analysis of the cosmological observations that only one sterile
neutrino type should exist is to some extent a model-dependent one, and
therefore for generality in this paper we consider a general model with three
sterile neutrinos. As noted above, the model is capable to reduce the number
of sterile neutrinos with main results being unchanged.

\section*{ACKNOWLEDGEMENTS}
\noindent
This study was carried out with the partial financial support of the Russian
Foundation for Basic Researches within the scientific project 14-22-03040
(ofi-m code).

\end{document}